


\magnification\magstep1
\parskip=\medskipamount
\hsize=6 truein
\vsize=8.2 truein
\hoffset=.2 truein
\voffset=0.4truein
\baselineskip=14pt
\tolerance=500


\font\titlefont=cmbx12
\font\abstractfont=cmr10 at 10 truept
\font\authorfont=cmcsc10
\font\addressfont=cmsl10 at 10 truept
\font\smallbf=cmbx10 at 10 truept
\font\bigmath=cmsy10 scaled \magstep 4


\def\bigtimes{\mathop{\vphantom{\sum}%
                      \lower2.5pt\hbox{\bigmath\char2}}}


\def\mapright#1{\smash{\mathop{\longrightarrow}\limits^{#1}}}
\def\maprights#1{\smash{\mathop{\rightarrow}\limits^{#1}}}
\def\maprightss#1{\mathop{\!\!\rightarrow\!\!}\limits^{#1}}
\def\mapdown#1{\Big\downarrow
  \rlap{$\vcenter{\hbox{$\scriptstyle#1$}}$}}


\def\diag{
\def\normalbaselines{\baselineskip2pt \lineskip3pt
\lineskiplimit3pt}
\matrix}


\newdimen\itemindent \itemindent=13pt
\def\textindent#1{\parindent=\itemindent\let\par=\resetpar%
\indent\llap{#1\enspace}\ignorespaces}

\let\oldpar=\par
\def\resetpar{\oldpar\parindent=0pt\let\par=\oldpar}

\font\ninerm=cmr9 \font\ninesy=cmsy9
\font\eightrm=cmr8 \font\sixrm=cmr6
\font\eighti=cmmi8 \font\sixi=cmmi6
\font\eightsy=cmsy8 \font\sixsy=cmsy6
\font\eightbf=cmbx8 \font\sixbf=cmbx6
\font\eightit=cmti8
\def\eightpoint{\def\rm{\fam0\eightrm}
  \textfont0=\eightrm \scriptfont0=\sixrm \scriptscriptfont0=\fiverm
  \textfont1=\eighti  \scriptfont1=\sixi  \scriptscriptfont1=\fivei
  \textfont2=\eightsy \scriptfont2=\sixsy \scriptscriptfont2=\fivesy
  \textfont3=\tenex   \scriptfont3=\tenex \scriptscriptfont3=\tenex
  \textfont\itfam=\eightit  \def\it{\fam\itfam\eightit}%
  \textfont\bffam=\eightbf  \scriptfont\bffam=\sixbf
  \scriptscriptfont\bffam=\fivebf  \def\bf{\fam\bffam\eightbf}%
  \normalbaselineskip=9pt
  \setbox\strutbox=\hbox{\vrule height7pt depth2pt width0pt}%
  \let\big=\eightbig \normalbaselines\rm}
\catcode`@=11 %
\def\eightbig#1{{\hbox{$\textfont0=\ninerm\textfont2=\ninesy
  \left#1\vbox to6.5pt{}\right.\n@space$}}}
\def\vfootnote#1{\insert\footins\bgroup\eightpoint
  \interlinepenalty=\interfootnotelinepenalty
  \splittopskip=\ht\strutbox %
  \splitmaxdepth=\dp\strutbox %
  \leftskip=0pt \rightskip=0pt \spaceskip=0pt \xspaceskip=0pt
  \textindent{#1}\footstrut\futurelet\next\fo@t}
\catcode`@=12 %


\outer\def\beginsection#1\par{\vskip0pt plus.2\vsize\penalty-150
\vskip0pt plus-.2\vsize\vskip1.2truecm\vskip\parskip
\message{#1}\leftline{\bf#1}\nobreak\smallskip\noindent}


\outer\def\subsection#1\par{\vskip0pt plus.2\vsize\penalty-80
\vskip0pt plus-.2\vsize\vskip0.8truecm\vskip\parskip
\message{#1}\leftline{\it#1}\nobreak\smallskip\noindent}


\newcount\notenumber

\def\note{\advance\notenumber by 1
\footnote{$^{\{\the \notenumber\}}$}}


\def\Sup{\hbox{${\cal S}(\Sigma)$}}
\def\S{\hbox{{$\Sigma$}}}
\def\SC{\hbox{${\bar{\Sigma}}$}}
\def\SP{\hbox{${\bar{P}}$}}
\def\Q{\hbox{${\cal Q}$}}
\def\QS{\hbox{${{\cal Q}(\Sigma)}$}}
\def\QRS{\hbox{${\cal Q}_R(\Sigma)$}}
\def\QRSC{\hbox{${\cal Q}_R(\bar{\Sigma})$}}
\def\RSC{\hbox{$Riem(\bar{\Sigma})$}}
\def\C{\hbox{$\cal C$}}


\rightline{Freiburg, THEP-92/32}
\bigskip
{\baselineskip=24 truept
\titlefont
\centerline{ON THE CONFIGURATION-SPACE TOPOLOGY}
\centerline{IN GENERAL RELATIVITY}
}

\vskip 1.1 truecm plus .3 truecm minus .2 truecm

\centerline{\authorfont Domenico Giulini\footnote*{
e-mail: giulini@sun1.ruf.uni-freiburg.de}}
\vskip 2 truemm
{\baselineskip=12truept
\addressfont
\centerline{Fakult\"at f\"ur Physik,
Universit\"at Freiburg}
\centerline{Hermann-Herder Strasse 3, D-W-7800 Freiburg, Germany}
}
\vskip 1.5 truecm plus .3 truecm minus .2 truecm

\centerline{\smallbf Abstract}
\vskip 1 truemm
{\baselineskip=12truept
\leftskip=3truepc
\rightskip=3truepc
\parindent=0pt

{\abstractfont
The configuration-space topology in canonical General Relativity
depends on the choice of the initial data 3-manifold. If the
latter is represented as a connected sum of prime 3-manifolds,
the topology receives contributions from all configuration spaces
associated to each individual prime factor. There are by now strong
results available concerning the diffeomorphism group of prime
3-manifolds which are exploited to examine the topology of the
configuration spaces in terms of their homotopy groups. We explicitly
show how to obtain these for the class of homogeneous spherical
primes, and communicate the results for all other known primes
except the non-sufficiently large ones of infinite fundamental group.
\par}}

\beginsection{Section 1. Introduction}

In recent years mathematicians have made progress in understanding
the diffeomorphism group of 3-dimensional manifolds. The object of
this paper is to show how this can be exploited to deepen our
{\sl topological} understanding of configuration spaces occurring
in pure General Relativity. In particular, we shall investigate
their homotopy groups and thus generalize already existing work on
the fundamental group [Wi]. Besides for its intrinsic interest, a
major motivation to study these topological structures steems from
the canonical quantisation programme for General Relativity. Here,
general arguments suggest a topological origin of certain interesting
features of quantum gravity (e.g. degenerate vacuum structure,
absence of anomalies, superselection sectors), resembling those
already familiar from other (successfully quantized) theories.
Certainly, the arguments given in the context of quantum gravity are
primarily meant to be of heuristic value, that is, they are believed
to really give insight into some aspects of quantum gravity by using
methods which are not necessarily believed to survive an eventual
rigorous formulation of it. Amongst others, there are two reasons
that entertain this belief: Firstly, arguments identical in structure
work in other field theories (e.g. Yang-Mills), where there is a
quantum theory, secondly, the arguments (properly formulated) rest
merely on general covariance and do not require more structural
details about quantum gravity.

In any generally covariant theory the topological structure of
configuration space receives characteristic imprints from the
diffeomorphism group, which is used to mutually identify physically
equivalent points on an auxillary space that labels physical states
in a redundant way. If this auxillary space is topologically trivial,
as it is the case in General Relativity, all the topological
information in the homotopy groups of the quotient is determined by
those of the diffeomorphism group. Generally, this holds whenever the
configuration space is given as the base of a principal fiber-bundle
with structure group the diffeomorphisms and contractible total
space, as will be explained below. In this case the topology of the
base is directly related to the topology of the fibres, and it is
their topology which we are going to investigate. In theories where
besides the diffeomorphisms there is an additional gauge group acting
(which also occurs in the ``connection'' formulation of General
Relativity [Ash]), additional topological structure is induced. In
these cases our analysis can be used to provide the diffeomorphism
contribution. In order to work within a fixed framework, we shall
argue within the standard framework of General Relativity. But, as
will become apparent, the investigation is really of a more general
kind.

In the sequel of this introductory section and Section 2 we shall
provide some basic material concerning the notion of configuration
spaces in General Relativity, 3-manifolds and their diffeomorphism
groups. In particular, the notion of a spinorial manifold is
introduced. A more technical point is deferred to Appendix 1. Proofs
of already existing results are only included when it seems
appropriate. Their setting given here might differ from the one
originally given. This sets the stage for the derivations of some new
results in Section 3. In Section 4 all the results next to some other
useful information is combined in a table, and some first
observations are made. This section should be accessible without
going through the main body of the paper. Appendix 2 combines into
five theorems some scattered results from the literature which we
made essential use of.

\subsection{Configuration Spaces, 3-Manifolds and Diffeomorphisms}

The specification of initial data in General Relativity starts with
the selection of a 3-manifold, \S,  on which initial data are constructed
in form of a Riemannian 3-metric and the extrinsic curvature. Together they
satisfy an elliptic system of four differential equations, the constraints,
which are separate from the evolution equations. As configuration
space we address the quotient-space obtained from the space of all 3-metrics
on $\S$, where those metrics which label the same physical sate are mutually
identified. This reduces three (the so-called momentum constraints, which are
linear in momenta) of the four constraint equations, the remaining one being
the so called Hamiltonian constraint (which is quadratic in the momenta).
The identification is generically given by the action of some normal
subgroup (possibly the whole group) of the diffeomorphism group, which we
choose to call its gauge part, since it connects redundant labels for the
same physical state.
General covariance then implies that the quotient of the full diffeomorphism
group with respect to the gauge part acts on the configuration space as
proper symmetries, which now connect different physical states. We shall
refer to those simply as symmetries. In case the gauge part exhausts all
of the diffeomorphism group, there will simply be no symmetries.

Exactly how much of all diffeomorphisms are considered to be gauge part
depends on the physical situation one likes to describe and cannot be
answered a priori within the formalism. In General Relativity two
major situations arise: Firstly, $\S$ is closed and represents the whole
universe, in which case there are no symmetries and the gauge part is the
whole diffeomorphism group. This situation we shall refer to as the closed
case. It is usually employed in classical- and quantum-cosmology.
Secondly, $\S$ represents an isolated part of the universe, so that
$\S$ is a manifold that outside some connected compact set is homeomorphic
to the complement of a closed ball in $R^3$, i.e., to the cylinder
$R\times S^2$. This cylinder can be thought of as the transition region
between the system under study and the ambient universe relative to which the
system is described. The gauge part is then given by those diffeomorphisms of
$\S$ that asymptotically die-off as one moves along the cylinder in an outward
direction. Slightly more precise, we may compactify $\S$ by a 2-sphere
boundary at the outer end of the cylinder and take the gauge part as those
diffeomorphisms that fix the boundary. They form a proper normal
subgroup within the full diffeomorphism group of $\S$. The action of
symmetries is then interpreted as changing the relative positions of the
system with respect to the ambient universe. This situation we shall refer to
as the open case.
\goodbreak

Mathematically the situations just described are surprisingly unique,
in the sense that essentially (i.e. up to discrete groups) no other
choice of a quotient symmetry group could have been made. This is due
to two facts. Firstly, that the connected component of the diffeomorphism
group of a closed manifold is simple, and, secondly, that the connected
component of the group of boundary-fixing diffeomorphisms of a manifold
with connected boundary is simple, and given by the unique non-trivial
normal subgroup of the connected component of all diffeomorphisms [McD].
For more than one boundary component (i.e. a $\S$ with more than one
asymtotic region, a case which we do not consider here), there will be
more normal subgroups according to those diffeomorphisms that fix only
some of the boundary components [McD]. For a closed manifold, a minimal
non-trivial normal subgroup of the diffeomorphism group is given by its
connected component, whereas in the open case a minimal choice is given
by the connected component of asymptotically trivial diffeomorphisms (larger
choices would be the connected component of all diffeomorphisms, or all
asymptotically trivial diffeomorphisms). In any case, the only way in
which continuous symmetry groups can arise is via asymptotic regions or
boundaries. Clearly, selecting minimal normal subgroups as gauge part
corresponds to a maximal choice of symmetries.

As it stands, the open and closed case do not seem to be intimately
related. We will argue, however, that, in a sense explained below, our
topological investigations cover both situations at the same time. Recall
that in the closed case the configuration space, as defined above, namely as
the quotient of the space of metrics modulo all diffeomorphisms, has a
non-trivial singularity structure (described in [Fi1]), due to the changing
dimensions of isotropy groups at metrics with different isometries.
Rather then working with the singular configuration space $\Sup$
(for superspace), where e.g. the global dynamics  is only defined by some
regular dynamics on a singularity free resulution space (reflection
conditions etc.), one may instead use the resolution space from the start.
First arguments as to why canonical quantum gravity should also be
formulated on a resolution space of superspace where already given in
[DeW]. It turns out that there is a canonical resolution space for $\Sup$
which we call $\QRS$. Its construction is explained in lucid detail in
[Fi2].

On the other hand, in the open case, an admissible and convenient way
for our topological investigations is to consider  the one-point
compactification $\SC:=\S\cup\{\infty\}$ of $\S$ by a point called
$\infty$, and then define the configuration space, $\QS$, by the space
of all metrics on $\SC$, $\RSC$, modulo the diffeomorphisms that fix the
frames at $\infty$. To answer topological questions we neither need to
specify fall-off conditions nor the precise functional space for the
metric. To start the construction, we fix an oriented frame $u$ at $\infty$.
A general linear transformation of the tangent space
$T_{\infty}(\SC)$ is said to be $\in SO(3)$, if its matrix representative
with respect to $u$ is $\in SO(3)$. Clearly, all conclusions to follow
are independent of the choice of $u$. Let us now define:
$$\eqalign{
D(\SC)         : &=\{\hbox{orientation preserving $C^{\infty}-$diffeomorphisms
                     on \SC}\}\cr
D_{\infty}(\SC): &=\{\phi\in D\ /\ \phi(\infty)=\infty\,,\quad
                                   \phi_*\vert_{\infty}\in SO(3) \}\cr
D_F(\SC)       : &=\{\phi\in D_{\infty}\ /\ \phi_*\vert_{\infty}=\hbox{id}\,.
                     \}\cr}
\eqno{(1.1)}
$$
Here, $D_{\infty}(\SC)$ represents those diffeomorphisms of $\S$ which
induce ``rigid'' roations on the 2-sphere at the end of the cylinder.
But since the space of orientation-preserving diffeomorphisms of the
2-sphere is homotopy equivalent to its isometries $SO(3)$ [Sm], we may for
our topological purposes represent $D(\S)$ by $D_{\infty}(\SC)$.
As an important example let us consider the particular class of open cases,
where the system under consideration is represented by asymptotically flat
metrics. Here, the restriction to $D_F$ as the gauge part is well motivated,
since we want to include configurations with non-vanishing angular momentum
at spatial infinity ($\infty$). These are included in $\RSC/D_F(\SC)$ but
not in the smaller space $\RSC/D_{\infty}(\SC)$. Diffeomorphisms on $\S$
must not disturb the fixed asymptotically euclidean structure, and are
therefore faithfully represented by $D_{\infty}(\SC)$. The symmetry group is
then given by $D_{\infty}(\SC)/D_F(\SC)$. However, motivated by its natural
appearence in canonical quantum gravity, where only the connected component
$D_F^0$ of $D_F$ appears as gauge part, we introduce the slightly larger
symmetry group ${\cal G}$ (note that by the discussion above this is the
maximal symmetry group):
$$
{\cal G}(\S):=D_{\infty}(\SC)/D^0_F(\SC)   \,.
\eqno{(1.2)}
$$
It is now true that \QS\ is the basis of a $D_F$- principal fibre bundle with
total space $\RSC$ [Bou][Fi2]:
$$
\diag{D_F(\SC)&\mapright{}&\RSC\cr
      &&\mapdown{\pi}\cr
      &&\QS\cr}
\eqno{(1.3)}
$$

We can now compare this to the closed case by considering the space $\QRSC$
for the closed manifold $\SC$. If we denote by $F(\SC)$ the bundle of
oriented frames over $\SC$, the resolved configuration space is defined as
the base of the following principal fibre bundle [Fi2]:
$$
\diag{D(\SC)&\mapright{}&\RSC\times F(\SC)\cr
      &&\mapdown{}\cr
      &&\QRSC\cr}
\eqno{(1.4)}
$$
Here, $D(\SC)$ acts on $F(\SC)$ by its standard lift, and the action is
free by the same argument as above. We now have the following result which
is part of Theorem 6.1 in [Fi2]:

\proclaim Theorem 1. The spaces $\QS$ and $\QRSC$ are diffeomorphic (as
ILH-manifolds).

For us this implies that we can focus attention to $\QS$. All the
statements we are going to make about the abstract topology of $\QS$
hold equally well for $\QRSC$. Keeping this in mind we shall never
mention $\QRSC$ again.

$\RSC$ is a convex open cone in the topological vector space of smooth
(0,2)-tensor fields. $D_F$ is topologized to make it a topological group
(i.e. at least as fine as compact open) with topological action
on \RSC. \Q\ is given the quotient topology which is the finest topology
for which $\pi$ is continuous (for more information on topologies of
mapping spaces see [Mi]). There are two important points to make.

1. The map $\pi$ is open and hence \Q's topology unique.

{\baselineskip=20truept{\eightpoint
\noindent Proof: take an open set ${\cal O}$ in $\RSC$.
Since $D_F$ acts as a topological group, its orbit, given by $\pi^{-1}(
\pi({\cal O}))$, is also open. Hence $\pi({\cal O})$ is an open set and
therefore $\pi$ an open map. Conversely, let $\pi$ be an open
map and $U\subset \QS$ an arbitrary set such that $\pi^{-1}(U)$ is
open. Then $U=\pi(\pi^{-1}(U))$ is open and therefore \QS's topology
equal or stronger than the quotient topology. But the quotient topology is
already the strongest one compatible with the continuity of $\pi$. Hence
it is unique. With respect to $Q$ it therefore makes sense to talk of
{\it its} topology.\par}}

2. Due to a theorem of Cerf's [C], the diffeomorphism and homeomorphism
groups of \SC\ are homotopy equivalent spaces. This allows us to be imprecise
about the function spaces we work in. In particular, it allows us
to use interchangeably $D_F(\SC)$ and $H_D(\SC)$, where the latter denotes the
space of homeomorphisms fixing a disc containing $\infty$ which is often
employed in the literature (e.g.[Fr-Wi],[He~-~L]).

{}From the contractibility of ${\rm Riem(\S)}$ and the homotopy exact
sequence associated with the bundle (2), we immediately obtain for all
$n\geq 0$:
$$
\pi_n(D_F(\SC))\cong \pi_{n+1}(\QS)\,.
\eqno{(1.5)}
$$
The investigation of the homotopy groups of \QS\ is thus reduced to
those of $D_F(\SC)$. No reference to the space of metrics
is made anymore. It has dropped out of the homotopy exact sequence due to its
contractibility ($\QS$ is a classifying space for the group $D_F$) and the
only topological features are those of $D_F$. This is why investigations
of this type bear a high degree of generality. For example, in so-called
higher derivative theories of gravity, $\RSC$ is replaced by $\RSC\times K$,
where $K$ is the linear (and hence contractible) space of sections in some
tensor bundle. The $D_F$ action is then still free and the total space still
contractible. The corresponding configuration spaces therefore still satisfy
(1.5).

Let us now decompose \SC\ into its prime factors $\SC_i$, explicitly
separating the irreducible primes $\SP_i$ from the non-irreducible
handles $S^2\times S^1$.

{\baselineskip=20truept{\eightpoint
\noindent
A prime $\SC_i$ is irreducible $\Leftrightarrow$ in $\SC_i$ every
2-sphere bounds a disc. For closed, orientable primes, the only
non-irreducible one is the handle.
\par}}
$$
\SC=\biguplus_{i=1}^n\SC_i=\left(\biguplus_{i=1}^l\SP_i\right)\uplus
                           \left(\biguplus_{i=l+1}^n S^2\times S^1\right)
                           \quad \SP_i\not\cong S^2\times S^1      \,,
\eqno{(1.6)}
$$
where we used $\uplus$ to denote the operation of taking the conencted sum.
It is known (e.g. [McCu]) that there is a homotopy equivalence ($\sim$)
$$
D_F(\SC)\sim \left(\bigtimes_{i=1}^l D_F(\SP_i)\right)\times
             \left(\bigtimes_{i=l+1}^n \Omega SO(3)\right)\times
             \Omega \C(\SC) \,,
\eqno{(1.7)}
$$
where the symbol $\Omega A$ stands for the loop-space of $A$, and  where
\C(\SC)\ is a space that (vaguely speaking) labels and topologizes the
relative configurations (sites) of the prime-manifolds $\SC_i$ in the
connected sum $\SC$.
In general its determination seems to be difficult and we refer to
[McCu][He-L] for more elaborate treatements. What interests us
here is that consequently we have the direct-product structure for the
homotopy groups $\pi_k$ ($k\geq 0$)
$$
\pi_k\left(D_F(\SC)\right)=
\left(\bigtimes_{i=1}^l \pi_k(D_F(\SP_i))\right)\times
\left(\bigtimes_{i=l+1}^n\pi_{k+1}(SO(3))\right)\times
\pi_{k+1}(\C)\,.
\eqno{(1.8)}
$$

This result rests on the restriction to the asymptotically trivial
diffeomorphisms $D_F$, for only in this case there is no topological
intertwinement of the diffeomorphisms with support inside the prime
factors $\SC_i$ (called internal diffeomorphisms) with those of general
support (called external diffeomorphisms). This would fail to hold if
one considered $D$ instead of $D_F$ as a simple counterexample shows
(see remark 1 in [He-McCu]). As far as the homotopy groups are concerned,
we now see how the determination of $\Q(\S)$'s topology is directly
related to the determination of those for its prime factors $\Q(\S_i)$.

\goodbreak
\beginsection{Section 2. Some preparatory material}

At the end of this section and the whole of Section 3 we will perform
explicit calculations for $\pi_0(D_F(\SC))$ and $\pi_k(D_F(\SC))$
($k\geq 1$) respectively, where $\SC$ is taken from the subclass
of homogeneous spherical primes. In order not to overload the actual
proofs, we shall establish some preparatory material first.

\subsection{A Closer Look at Diffeomorphisms}

The strategy for our calculations is very simple: there are
hard theorems available concerning the topological structure of $D(\SC)$
for $\SC$ prime. For convenience we collect some of them in Appendix 2.
Most interesting for us is the question of validity of a conjecture
made by Hatcher [Ha2] (abreviated HC, in the literature also refered
to as the generalized Smale conjecture). It asserts that for spherical
primes the spaces of diffeomorphisms and isometries are homotopy equivalent.
Restricted to the 3-sphere, this is known as the Smale conjecture which
has been shown to hold in [Ha4]. Some of the proofs presented in this paper
require the validity of (HC), the status of which has been indicated in the
table presented at the end.

{}From now on we shall sometimes drop the explicit reference to the manifold
$\SC$ by just writing $D_F$, $D_{\infty}$, etc., without any arument.
We obtain information about $D_F$ by relating it to $D$ by some standard
fibrations which we shall now describe.
First note that orientable 3-manifolds are always parallelizable.
Their bundle of oriented frames, $F(\SC)$, is thus topologically
the product $\SC\times GL^+(3,R)$. The different diffeomorphism groups in
(1.1) are topologically related by the following three principal fibre
bundles ($u$ still denotes the fixed frame at $\infty$):
$$
\diag{
D_F&\mapright{\hat i}&D&&\cr
&&\mapdown{\hat p}&&\,{\hat
p}(\phi):=T\phi(u)=(\phi(\infty),\phi_*|_{\infty}(u))\cr
&&F(\SC)&&\cr }
\eqno(2.1)
$$$$
\diag{
D_{\infty}&\mapright{i}&D&&\cr
&&\mapdown{p}&&\quad p(\phi):=\phi(\infty)\qquad\qquad\qquad\quad\cr
&&\ \SC\ &&\cr }
\eqno(2.2)
$$$$
\diag{
D_F&\mapright{{\tilde i}}&D_{\infty}&&\cr
&&\mapdown{\tilde p}&&\quad\!\!{\tilde p}(\phi):=\phi_*|_{\infty}
\qquad\qquad\qquad \quad\cr
&&SO(3)&&\cr }
\eqno(2.3)
$$

As a first application we introduce the concept of spinoriality of a manifold
$\SC$. Associated with (2.3) is the fibration
$$
\diag{D_F(\SC)/D^0_F(\SC)&\mapright{{\tilde i}}&{\cal G}(\S)\cr
      &&\mapdown{{\tilde p}}\cr
      &&SO(3)\cr}
\eqno{(2.4)}
$$
where ${\cal G(\S)}$ is the symmetry group defined in (1.2). From the
associated exact homotopy sequence we infer
$$\diag{
0 & \maprights{}        &  \pi_1({\cal G})       & \maprights{{\tilde p}_*}
  & Z_2 & \maprights{}  &  \pi_0(D_F)            & \maprights{}
  & \pi_0({\cal G})     &  \maprights{} & 0\cr}
\eqno{(2.5)}
$$
which, by injectivity of ${\tilde p}_*$, tells us that there are only
two possibilities:

1. $\pi_1({\cal G})=0$ hence $\pi_0(D_F)$ is a $Z_2$ extension of
$\pi_0({\cal G})$. In this case (2.4) is non-trivial and ${\cal G}$
is given by $\{\pi_0(D_F)\times SU(2)\}/Z_2$. Here the $Z_2$ is
generated by $(-1,-1)$ where the $-1$ in the left factor generates
some $Z_2$ in $\pi_0(D_F)$ and the $-1$ in the right factor generates
the centre of $SU(2)$.

2. $\pi_1({\cal G})=Z_2$ and $\pi_0(D_F)=\pi_0({\cal G})$,
in which case (2.3) is trivial and ${\cal G}(\S)$ a direct product
$\pi_0(D_F)\times SO(3)$.

In the first case, ${\cal G(\S)}$ contains $SU(2)$ but not $SO(3)$ as a
subgroup. Let a manifold, $\S$, for which this is the case, be for obvious
reasons called {\sl spinorial}. Whether a manifold is spinorial is a
purely topological question and has been decided for all known prime
manifolds (see Theorem 2 below). In view of Corollary 1 (below) this
provides sufficient information for the general case. From (2.3) one
obtains
$$\diag{
0&\maprightss{}&\pi_1(D_F)&\maprightss{}&\pi_1(D_{\infty})&
\maprightss{{\tilde p}_*}&Z_2&\maprightss{\partial_*}&\pi_0(D_F)&
\maprightss{}&\pi_0(D_{\infty})&\maprightss{}&1  \cr}
\eqno{(2.6)}
$$
$$\eqalign{
\hbox{    \S\ is spinorial} & \Leftrightarrow\partial_*\,\hbox{into}
                              \Leftrightarrow \pi_1(D_F(\SC))\cong
                                              \pi_1(D_{\infty}(\SC)) \cr
\hbox{\S\ is not spinorial} & \Leftrightarrow {\tilde p}_*\,\hbox{onto}
                              \Leftrightarrow \pi_0(D_F(\SC))\cong
                                              \pi_0(D_{\infty}(\SC)) \cr}
\eqno{(2.7)}
$$

For later application let us cite the following known results in form of two
lemmas, a theorem and a corollary.

\proclaim Lemma 1. If \SC\ is an irreducible prime manifold (i.e. a prime
different from $S^1\times S^2$), then the bundle projection $p$ of (2.2)
has the property that $p_*$ maps $\pi_1(D)$ onto the centre $C$ of
$\pi_1(\SC)$.\par

\noindent{\bf Proof.} That $p_*(\pi_1(D(\SC)))\subset C$ follows
from Corollary 5.22 of ref. [McCa] for arbitrary $\SC$ (his/her remark
5.24). Surjectivity onto $C$ was shown in section V of ref. [Wi] under the
hypothesis that homotopy implies isotopy of diffeomorphisms. According to
Theorem A1 in Appendix 2 this is now known to hold for all primes in our
table~$\bullet$\par

\proclaim Lemma 2. If \SC\ is not spinorial then the bundle projection
$\hat p$ of (2.1) has the property that ${\hat p}_*(\pi_1(D))$ contains
$(1,-1)\in\pi_1(F(\SC))\cong \pi_1(\SC)\times Z_2$, where $1$ is the identity
in $\pi_1(\SC)$ and $-1$ the generator of $Z_2$.\par

\noindent{\bf Proof.} This is proven in Lemma 2.1 of ref. [Fr-Wi].
The simple and instructive proof is worth a look at this point.
Let $s\mapsto R_s\in D^0_{\infty}$ (the connected component of $D_{\infty}$)
be the rotation of a 3-disc $D_1$ aginst a slightly larger, concentric
2-sphere $S_2$ (compare Appendix~1). Non-spinoriality implies the existence
of a path $s\mapsto \phi_s\in D^0_F$ so that $\phi_0=R_1$ and $\phi_1=id$.
The product path, $\gamma$, defined by
$$
\gamma:=\cases{R_{2s}        & for $s\in [0,{1\over 2}]$\cr
               \phi_{(2s-1)} & for $s\in [{1\over 2},1]$\cr}
$$
defines a loop at $id$ in $D^0_{\infty}$ satisfying
${\tilde p}_*([\gamma])=-1$. Moreover, from the end of the exact sequence
for (2.2) (shown in (3.3)), we have $p_*([\gamma])=p_*\circ i_*([\gamma])=1$,
so that ${\hat p}_*([\gamma])=(p_*([\gamma]),{\tilde p}_*([\gamma]))=(1,-1)
\in \pi_1(\SC)\times Z_2$~$\bullet$

\proclaim Theorem 2. The only non spinorial prime manifolds are the
lens spaces, $L(p,q)$, and the handle $S^1\times S^2$.\par

\noindent{\bf Proof.} Non-spinoriality for $L(p,q)$ and $S^1\times S^2$
can actually be visualized. The demonstration of this fact is deferred to
the Appendix 1. In ref. [He], chapter 4.3, Theorem 1 implies
spinoriality for the following prime manifolds: Those with infinite
fundamental group different from $S^1\times S^2$ (the so-called $K(\pi,1)$'s),
and those with finite fundamental group which have a non-cyclic 2-Sylow
subgroup. The remaining cases consist of some $S^3/G$ with non cyclic $G$
for which HC is known to hold true. Using this fact, they where shown to
be spinorial in ref. [Fr-Wi], Theorem 2.2 and the remark
following Corollary 2.2. For the latter one needs to add that in the
meantime the validity of HC for the spaces $S^3/D^*_{4m}\,m\geq 2$ has been
shown in ref. [McCu-R].

{\baselineskip=20truept{\eightpoint
Due to our strategy to be explicit in the case of homogeneous $S^3/G$'s,
and anticipating some results from the next subsection,
we want to give a simple proof of spinoriality for non-cyclic $G$ under
the hypothesis of validity of HC. (More precisely, it only relies on one of
its implications, namely that $\pi_1(D)$ and $\pi_1(Isom)$
have the same number of generators. We are, however, not aware of a single
case where this but not the full HC is known to hold.) By discreteness of
$S(G)$ we learn from (2.12) that ${\bar p}_*$ maps $Z_2\cong \pi_1(Isom(G))$
injectively into $G=\pi_1(S^3/G)$. Since $\bar p$ is just the restriction of
$p$ in (2.2), which also appears as the first component of $\hat p$ in (2.1),
we learn that ${\hat p}_*(\pi_1(D))$ contains the element $(-1,1)$ or
$(-1,-1)$ [by Lemma 1, ${\hat p}_*$ maps $\pi_1(D)$ into
$Z_2\times Z_2=centre\,G\times\pi_1(SO(3))]$. By HC,
$\pi_1(D)\cong\pi_1(Isom)= Z_2$ so that ${\hat p}_*(\pi_1(D))$ cannot in
addition contain $(1,-1)$ by mapping only one generator. Lemma 2 then
implies spinoriality $\bullet$\par}}

\proclaim Corollary 1.
A closed, oriented, connected 3-manifold $\SC$ is
non-spinorial, if and only if its prime decomposition
consists entirely of lens spaces and handles.

\noindent{\bf Proof.}
In view of Theorem 2 we need to prove that such a manifold
is non-spinorial, if and only if none of its prime factors
is spinorial. For this we assume that $\SC$ is the connected
sum of $n$ prime factors which build up $\SC$ as follows
(we follow ref. [McCu]): Take a 3-sphere and on it n+1 closed,
disjoint 3-discs $D_i$, $0\leq i \leq n$,
with 2-sphere boundaries $\partial D_i=S_i$. We take $D_0$ as a
neighbourhood of $\infty$, remove the interiors of $B_i$
for $i\geq 1$, and construct $\SC$ by identifying the $n$ 2-sphere
boundaries $S_i$ ($1\leq i\leq n$) with the corresponding
boundaries that one obtains by removing an open 3-disk on
each prime factor respectively. In $\SC$ we now connect $S_i$ with
$S_{i+1}$ for each $1\leq i \leq n-1$ by a thin cylindrical tube
(topology $I\times S^1$) and obtain a single new 2-sphere, $S$ (the
connected sum of all $S_i$ for $i\geq 1$), which is isotopic
to $S_0$. A $2\pi$-rotation parallel to $S_0$ (i.e. a rotation
parallel to $S_0$ and an arbitrarily close concentric one, as
explained in Appendix 1) is also isotopic to such a rotation parallel
to $S$. The latter one may be choosen to have the
connecting tubes as axis (i.e. the rotation acts only on the
$S^1$ part of $I\times S^1$). Shrinking the tubes to zero diameter
defines an isotopy to rotations parallel to each $S_i$ for
$1\leq i \leq n$ (compare the remark after Theorem 1, chapter 4.3, of
reference [He]). From (1.8) for $k=0$ and Theorem 2 we infer that this
diffeomorphism corresponds to the identity element of $\pi_0(D_F(\SC))$,
if and only if the connected sum does not contain a single spinorial
prime~$\bullet$

Corollary 1 suggests that non-spinorial manifolds are somewhat more
special than spinorial ones. Note, however, that the non-spinorial primes
still suffice to build up (in a non-unique fashion) manifolds of any given
homology by taking connected sums. Non-spinorial manifolds are, therefore,
not {\sl homologically} special. Note also that the lens spaces $L(p,q)$ are
homogeneous, if and only if $q=\pm 1\,\hbox{mod}\,p$. These are therefore
the only non-spinorial primes in the class of homogeneous space forms to
which we specialize in the following subsection. Finally, we
note that the existence of spinorial manifolds has been already used in the
literature to speculate about the existence about certain spinorial states
in quantum gravity [Fr-So]. We will come back to this point in Observation 1
of Section~4.

\subsection{Homogeneous Spherical Primes}

In this subsection we consider a special class of elliptic spaces, that is,
spaces of the form $S^3/G$, where
$G$ is a finite subgroup of $SO(4)$ with free action on $S^3$. We identify
$S^3$ with the group manifold of $SU(2)$ with its standard bi-invariant (round)
metric and use the isomorphism $SO(4)\cong (SU(2)\times SU(2))/Z_2$, where
the $Z_2$ is generated by $(-1,-1)$. Elements of $SO(4)$ are then written as
$Z_2$-equivalence classes $[g,h]$ with $g,h\in SU(2)$, and elements of $S^3/G$
as $G$-equivalence classes $[x]$ with $x\in S^3$.
We can now write down the (left) $SO(4)$-action on $S^3$
$$
SO(4)\times S^3\rightarrow S^3 \,,\qquad
[g,h]\times x \mapsto g\cdot x\cdot h^{-1}\,.
\eqno{(2.8)}
$$
$SO(4)$ has a left- and right- $SU(2)$ subgroup, given by the sets
$SU(2)_L:=[SU(2),1]$ and $SU(2)_R:=[1,SU(2)]$ respectively.
It also has an obvious diagonal- $SO(3)$ subgroup given by all elements
of the form $[g,g]$. We call it $SO(3)_D$. If $N_{SO(4)}(G)$ denotes the
normalizer of $G$ in $SO(4)$, it is easy to see that the residual isometry
group acting on $S^3/G$ is given by $Isom(S^3/G)=N_{SO(4)}(G)/G$:
$$
N_{SO(4)}(G)\times S^3/G\rightarrow S^3/G\,,\qquad
[g,h]\times [x] \mapsto [g\cdot x\cdot h^{-1}]\,.
\eqno{(2.9)}
$$
It acts transitively on $S^3/G$, if and only if $G$ is a finite, freely acting
subgroup of either $SU(2)_L$ or $SU(2)_R$. As subgroups of $SU(2)$ these are
given by $Z_p$, $D^*_{4n}$ for $n\geq 2$, $T^*$, $O^*$ and $I^*$ which denote
the cyclic group of order $p$ and the $SU(2)$-double covers (denoted by *)
of the symmtery groups of the n-prism, tetrahedron, octahedron and icosahedron
of orders $4n$, 24, 48 and 120 respectively.
We shall now restrict to theses cases and may choose $G=G_R$, that is, we
may choose $G$ to sit inside $SU(2)_R$, for the manifold obtained by using
$SU(2)_L$ would be certainly homeomorhic.

{\baselineskip=20truept{\eightpoint
For our topological purposes we may indeed restrict ourselves
to the right action for identifying $S^3$ to $S^3/G$.
But as prime manifolds in the oriented category one needs to
distinguish those obtained by right and left identifications.
This follows from the fact that a homogeneous $S^3/G$ does not
allow for any orientation reversing diffeomorphism if $G\not=Z_i$
for $i\leq 2$ [Wi] and uniqueness of the prime decomposition in the
oriented category. For those $G$ this implies that
$S^3/G\uplus S^3/G$ is not homeomorphic to (the minus sign
indicates reversed orientation) $S^3/G \uplus (-S^3/G)$.
On the other hand, the difeomorphism $\phi$ that relates the two
manifolds obtained by right and left identifications is just given
by $[p]_R\mapsto \phi([p]_R):=[p^{-1}]_L$, where $p\in S^3\cong
SU(2)$ and $[\cdot]_R, [\cdot]_L$ denote the right and left cosets
respectively. But this is an orientation reversing diffeomorphism.\par}}

For all but the cyclic groups $Z_p$ of odd order, $G$ contains the centre
$Z_2$ of $SU(2)$. Standard properties of groups and their quotients imply the
following homomorphism equivalences:
$$
Isom(S^3/G)\cong SO(3)\times{N_{SU(2)}(G)\over G}\cong
                 SO(3)\times{N_{SO(3)}(G/Z_2)\over G/Z_2}\,.
\eqno{(2.10)}
$$

Let $S(G)$ denote the (unique) conjugacy class of stabilizer subgroups for
$Isom(G)$'s action on $S^3/G$. We shall usually identify it with the
stabilizer subgroup at $[e]\in S^3/G$,
where $e$ is the identity element of $S^3\cong SU(2)$. One easily shows that
$S(G)\cong N_{SU(2)}(G)/Z_2\subset SO(3)_D$, where $Z_2$ is the
centre of $SU(2)$. For $G\not =Z_p$, where $p$ odd, this can be written as
$S(G)\cong N_{SO(3)}(G/Z_2)$. As a closed subgroup of $SO(3)$ it must be
either discrete, or a finite disjoint union of circles, or the whole of
$SO(3)$. The last case is realized for $G=Z_2$ and the second case for
$G=Z_p$ ($p>2$), where $S(G)$ is given by two circles in two perpendicular
planes (viewed as subset of unit quaternions) one of which is the unique
1-parameter subgroup containing $Z_p$. In the first case we have explicitly
(note: $D_{4n}$ is the projection of $D^*_{8n}$ into $SO(3)$):
$$
S(G)=\cases{O      & for $G=D^*_8, T^*$ and $O^*$   \cr
            D_{4n} & for $G=D^*_{4n}$               \cr
            I      & for $G=I^*$                    \cr}
\eqno{(2.11)}
$$

By restricting the total space of the bundle (2.2) to the isometries,
and, accordingly, the fibre $D_{\infty}$ to $S(G)$, we obtain the
principal-bundle
$$
\diag{
S(G) & \mapright{i} & Isom(S^3/G) && \cr
&&\mapdown{\bar p}&&\quad {\bar p}([g,h]):=[g\cdot x\cdot h^{-1}]\,,
\qquad\qquad\quad\cr
&&\ S^3/G &&\cr }
\eqno{(2.12)}
$$
where $[g,h]\in N_{SO(4)}(G)$ and $x\in S^3$ represents the point $\infty=[x]$
on $S^3/G$. The projection $\bar p$ is just the restriction of $p$ in (2.2)
to $Isom(S^3/G)$.
Next we prove a lemma which we shall use for the calculation of
$\pi_k(D_F)$, $k\geq 2$, later on.

\proclaim Lemma 3. If $G\not =Z_2$ is a non-trivial finite subgroup of $SU(2)$
which acts freely on $S^3$ via the standard orthogonal action, and if HC holds
for $S^3/G$, then the projection map $p$ from bundle (2.2) induces isomorphisms
$$
p_*\,:\ \pi_k(D(S^3/G))\longrightarrow\pi_k(S^3/G)\qquad\forall k\geq 2\,.
$$

\noindent{\bf Proof.} The validity of HC implies that the
inclusion $I:\,Isom(S^3/G)\hookrightarrow D(S^3/G)$ induces isomorphisms
$I_*:\,\pi_k(Isom(S^3/G))\rightarrow\pi_k(D(S^3/G))\,\forall k$. Since for
$G\not =Z_2$ we have $\pi_k(S(G))=0\,\forall k\geq 2$, the homotopy exact
sequence for (2.12) implies that
${\bar p}_*:\pi_k(Isom(S^3/G))\rightarrow\pi_k(S^3/G)$
is an isomorphism $\forall k\geq 2$. On the other hand, $\bar p=p\circ I$,
and hence ${\bar p}_*=p_*\circ I_*$, which proves the claim $\bullet$\par

\subsection{Calculation of $\pi_0(D_F(S^3/G))$.}

For the spherical primes $S^3/G$, the zeroth homotopy group of $D_F$ has
already been calculated in [Wi]. Here we give a separate calculation in
the homogeneous subclass.

We start with the lens spaces $L(p,1)$. Lemma 1 applied to the
exact homotopy sequence for (2.2) implies that its very last two
group entries are isomorphic: $\pi_0(D_{\infty})\cong \pi_0(D)$.
Theorem 2 and (2.7) then imply $\pi_0(D_F)\cong\pi_1(D)\cong \pi_0(Isom)$
(Theorem A2, Appendix 2). This is isomorphic to $Z_2$ for $p>2$, and to
the trivial group for $p\leq 2$.

For $G$ non-cyclic, we have that $Z_2:=\{1,-1\}\subset SU(2)$ is the centre
and the centralizer of $G$ in $SU(2)$. This implies that the action of
$S(G)$ on $S^3/G$, $[p]\mapsto [gpg^{-1}]$, is effective on the coset
$[e]\subset S^3$ ($e$=identity in $SU(2)\cong S^3$) and therefore on the
fundamental group $\pi_1(S^3/G,[e])$. Thus no nontrivial element in $S(G)$
lies in the identity component in $D_{\infty}$ (we take $\infty=[e]$), so that
$\pi_0(D_{\infty})$ contains the subgroup $S(G)$. We wish to show that it
actually saturates all of $\pi_0(D_{\infty})$. This can be done if
$\pi_0(D)\cong\pi_0(Isom)$ (see Theorem A2), for then we infer from Lemma~1
that the last part of the sequence for (2.2) just says that
$\pi_0(D_{\infty})$ is a $G/Z_2$ - extension of $\pi_0(D)$, and therefore
($\vert\cdot\vert$ denotes the order of the group $\cdot$):
$$
\vert\pi_0(D_{\infty})\vert={\vert G\vert\over 2}\vert\pi_0(D)\vert=
{\vert G\vert\over 2}{\vert N_{SO(3)}(G/Z_2)\vert\over\vert G/Z_2\vert}=
\big\vert N_{SO(3)}(G/Z_2)\big\vert\,.
\eqno{(2.13)}
$$
Next, by spinoriality (Theorem 1), we know from (2.6) and (2.7)
that $\pi_0(D_F)$ is a $Z_2$ - extension of $D_{\infty}$. We wish to
show that it must be the $SU(2)$-double cover $N_{SU(2)}(G)$. To do this
in detail requires some notation.

Let $N_{SU(2)}(G)=\{g_1,\cdots , g_N\}$, $G=\{g_1,\cdots ,g_K\},\, K\leq N$,
where $g_1=e$, and $\{\theta_1,\cdots ,\theta_N\}\subset su(2)$ (the
Lie-algebra of $SU(2)$) such that $\exp(\theta_i)=g_i$ (we take indices
$i,\cdots$ to run from 1 to $N$, and $a,\cdots$ from 1 to $K$). There is a
homomorphism $\sigma:\, N_{SU(2)}(G)\rightarrow P_K$ into the permutation
group of $K$ objects, $g_i\mapsto \sigma(g_i)=:\sigma_i$, defined by
$g_a g^{-1}_i= g^{-1}_i g_{\sigma_i(a)}$. Let $r:\, S^3\rightarrow R$ be the
distance function from $e$ (with respect to the bi-invariant metric) and
$\rho$ a $C^{\infty}$ step-function, such that $\rho(r)=0$ for $r\leq \epsilon$
and $=1$ for $r\geq 2\epsilon$. We then define $\lambda:=\rho\circ r$. Further,
we let $B^2_1$ and $B^1_1$ be the closed $2\epsilon$- and $\epsilon$- balls
around $e=g_1$ respectively. We right-translate them to $2\epsilon$-
respectively $\epsilon$- balls centred at $g_a$, i.e., $B^2_a=B^2_1\cdot g_a$
and $B^1_a= B^1_1\cdot g_a$ for each $a$. We further take $\epsilon$ small
enough for $B^2_a$ to lie within a regular neighbourhood with respect to the
covering $S^3\rightarrow S^3/G$. In particular, they are all disjoint. Their
projections into $S^3/G$ are called $B^2$ and $B^1$.

A diffeomorphic action of $N_{SU(2)}(G)$ on $S^3-\cup_{a=2}^K B^2_a$ is then
defined via
$$
T_i:\quad p\mapsto\exp(\lambda(p)\theta_i)\cdot p\cdot
                  \exp(-\lambda(p)\theta_i)\,.
$$
It is easy to see that:
(i)   $T_i$ leaves $B^1_1$ pointwise fixed,
(ii)  $T_i$ leaves $B^2_1-B^1_1$ invariant,
(iii) $T_i$ leaves $S^3-\cup_{a=1}^K B^2_a$ invariant and
(iv)  $T_i$ maps $\partial B^2_a$ onto $\partial B^2_{\sigma_i(a)}$.
In fact, any point in a small closed outer collar-neighbourhood of $\partial
B^2_a$ can be written as $pg_a$, where $p$ is from such a neighbourhood of
$\partial B^2_1$. Under $T_i$ it is mapped according to
$$
pg_a\mapsto g_ipg_ag^{-1}_i=T_i(p)g_{\sigma_i(a)}\,.
$$
We now use this formula to smoothly extend the map to all of $S^3$, that is,
for any point $pg_a\in B^2_a$, where $p\in B^2_1$, we set
$$
T_i(p g_a)=T_i(p) g_{\sigma_i(a)} \,,
$$
where $T_i(p)$ is defined above. By construction, this action projects onto
$S^3/G$ and acts on the fundamental group of $S^3/G$ by simple conjugation:
$p\mapsto g_ipg^{-1}_i$; for, as before, a path on
$S^3$ from $e$ to the $g_a$ is mapped to a path from $e$ to $g_{\sigma_i(a)}$.
Only the centre $Z_2$ of $N_{SU(2)}(G)$ acts trivially on the fundamental
group. However, the action of this $Z_2$ just corresponds to a relative
$2\pi$-rotation of the spheres $\partial B^2_1$ and $\partial B^1_1$,
which, by spinoriality, is not isotopic to the identity keeping a frame
at $[e]$ fixed. We have thus shown that $\pi_0(D_F(S^3/G))$ contains, and is
hence equal to $N_{SU(2)}(G)$.

\beginsection{Section 3. Calculation of $\pi_k(D_F(S^3/G))$ for $k\geq 1$}

Within the class of spaces considered here, $\SC=S^3$ ($G=Z_1$) and
$\SC=RP^3$ ($G=Z_2$) receive a spacial status due to them being group
manifolds. This allows us to make a somewhat more concise statement than
in the other cases. We have the following
\medskip
\halign{\bf#&\sl#\cr
Lemma 4. &a): If $\SC$ is a topological group, (2.2) is
trivial.\cr
&b): For $\SC\cong RP^3$ or $\,\cong S^3$, (2.3) is trivial.\cr}
\par\nobreak\medskip
\noindent
{\bf Proof.}$\,$  a): Define a global section $\sigma:\,\S\rightarrow D$,
$x\mapsto L_x$ (=left translations). We have $p\circ \sigma ={\rm id}|_{\SC}$.
A global trivialisation is given by \nobreak

$$\eqalign{
\phi^{-1}&:\;\SC\times D_{\infty}\rightarrow D\,;\quad (x,h)\mapsto L_x
\circ h\cr
\phi &:\;D\rightarrow \SC\times D_{\infty}\,;\quad g\mapsto
\left(p(g),L^{-1}_{p(g)}
\circ g\right)\cr }
$$
b): Define a global section $\sigma:\,SO(3)\rightarrow D_{\infty}$, $\alpha
\mapsto {\rm Ad}(\alpha)$, where ${\rm Ad}(\alpha)(p)=\alpha
p\alpha^{-1}\,\forall p\in \SC$. We have that ${\rm Ad(\alpha)}_*|_e= \alpha$,
and hence ${\tilde p}\circ \sigma={id}|_{SO(3)}$. A corresponding
trivialisation is given by \par \nobreak
$$\eqalign{
\phi^{-1}&:\;SO(3)\times D_F\rightarrow D_{\infty}\,;\quad (\alpha,h)\mapsto
{\rm Ad}(\alpha)\circ h\cr
\phi&:\;D_{\infty}\rightarrow SO(3)\times D_F\,;\quad g\mapsto\left({\tilde p}
(g),{\rm Ad}^{-1}({\tilde p}\circ g)\right)\;\bullet\cr}
$$

By HC we know that $D(S^3)$ and $D(RP^3)$ have the homotopy type of
$SU(2)\times SO(3)$ and $SO(3)\times SO(3)$ respectively. Hence we
have the
\proclaim Theorem 2. \Q(\S) has alltogether trivial homotopy groups for
$\S=S^3$ or $\S=RP^3$. \par\noindent
{\bf Proof:} By the previous lemma, the bundles (2.2) and (2.3) are product
bundles. We thus have for all $k\geq 0$:
$$\eqalignno{
         \pi_k(D) & \cong\pi_k(\SC)\times\pi_k(D_{\infty})   &(3.1)  \cr
\pi_k(D_{\infty}) & \cong\pi_k(D_F) \times\pi_k(RP^3)        &(3.2)  \cr}
$$
Setting \SC\ either equal to $S^3$ or $RP^3$, and inserting the space
$Isom$ for $D$ proves the claim $\bullet$ \par

\subsection{Calculation of $\pi_1(D_F(\SC))$}

Since for $\SC=S^3/G$ we have $\pi_2(\SC)=0$, the exact sequences for (2.2)
and (2.1) end as follows:
$$\eqalignno{
&\diag{
0&\maprightss{}&\pi_1(D_{\infty})&\maprights{i_*}&\pi_1(D)&\maprights{p_*}&G&
\maprights{}&\pi_0(D_{\infty})&\maprights{}&\pi_0(D)&\maprights{}&1\cr}\quad
&(3.3) \cr
&\diag{
0&\maprightss{}&\pi_1(D_F)&\maprightss{{\hat i}_*}&\pi_1(D)&
\maprightss{{\hat p}_*}&
G\times Z_2&\maprightss{}&\pi_0(D_F)&\maprightss{}&\pi_0(D)&\maprightss{}&1\cr}
\quad&(3.4) \cr}
$$

We shall first deal with $G$ non-cyclic. For non-cyclic $G$ we have
$centre G=Z_2$ and by $HC$ $\pi_1(D)=\pi_1(Isom)=Z_2$, so that Lemma 1
tells us that $p_*$ in (3.3) is an isomorphism, and, therefore,
$\pi_1(D_{\infty})=0$. Spinoriality together with formulae (2.7) then imply
$\pi_1(D_F)=0$.

For $G=Z_p$, $\SC\cong L(p,1)$ is non-spinorial so that the Lemmas 1 and 2
imply ($G=centre G$) that ${\hat p}_*$ in (3.4) is onto. HC then implies that
(3.4) reduces to
$$
\diag{
0&\mapright{}&\pi_1(D_F)&\mapright{}&Z\times Z_2&\mapright{{\hat p}_*}&
Z_p\times Z_2&\mapright{}&0\,,\cr}
\eqno{(3.5)}
$$
where $\pi_1(D_F)\cong ker\,{\hat p}_*$. For even $p$ it is immediate that
$ker\,{\hat p}_*$ cannot contain the $Z_2$ (within $Z\times Z_2$), for,
otherwise, ${\hat p}_*$ cannot be onto. For $p$ odd we argue as
follows: $N_{SU(2)}(Z_p)$ contains the (unique) circle group in which
$Z_p$ lies. We consider the loop in $SO(3)_D$ obtained by tracing through
half that circle in the right- and left- $SU(2)$ at the same time.
This loop is easily seen to generate $(p,-1)\in Z\times Z_2\cong\pi_1(D)$,
and, since it fixes $[e]$ and induces a $2\pi$-rotation in the tangent space,
has an image $(a,-1)\in Z_p\times Z_2$ under ${\hat p}_*$ for some $a$.
Hence, since $p$ is odd, $(1,-1)\in Z\times Z_2$ cannot be in the kernel of
${\hat p}_*$. We have thus shown that $Z_2\not\in ker\,{\hat p}_*$ so that
$\pi_1(D_F)\cong Z$.

\subsection{Calculation of $\pi_k(D_F)$ for $k\geq 2$}

A typical piece of the exact sequence for (2.2) looks like:
$$\diag{
\cdots&\maprightss{}&\pi_{k+1}(D)&\maprightss{p_*^{(k+1)}}&\pi_{k+1}(\SC)&
\maprightss{}&\pi_k(D_{\infty})&\maprightss{}&\pi_k(D)&\maprightss{p_*^{(k)}}&
\pi_k(\SC)&\maprightss{}&\cdots\cr}
\eqno{(3.6)}
$$
By Lemma 3, $p_*^{(k)}$ is an isomorphism for $k\geq 2$, so that
$\pi_k(D_{\infty})=0\quad\forall k\geq 2$. The exact sequence for (2.3) then
implies $\pi_k(D_F)\cong \pi_{k+1}(SO(3))\quad\forall k\geq 2$.

\subsection{Summary}

According to (1.5) we can summarize the homotopy groups of the (connected)
configuration space $\Q(S^3/G)$ for $S^3/G$ homogeneous:
$$\eqalignno{
\pi_k(\Q(RP^3))&\cong 0\quad\forall k\geq 1 &(3.7)\cr &\cr
\pi_k(\Q(L(p,1)))&\cong\cases{Z_2        & for $k=1$\cr
                              Z          & for $k=2$\cr
                              \pi_k(S^3) & for $k\geq 3$\cr} &(3.8)\cr &\cr
\pi_k(\Q(S^3/G))&\cong \pi_k\left(S^3/N_{SU(2)}(G)\right)\quad\forall k\geq 1,
\quad\hbox{and $G$ non-cyclic}
&(3.9)\cr}
$$
In particular the last equation is very suggestive. We stress, however,
that we did not establish a homotopy equivalence
$\Q(S^3/G)\sim S^3/N_{SU(2)}(G)$.

\beginsection{Section 4. Results and some Observations}

$${\eightpoint
\matrix{
\hbox{Prime \SC}&&\quad HC\quad &&\quad \pi_1(\Q(\S))\quad &&\quad
\pi_2(\Q(\S)) \quad &&\pi_k(\Q(\S))\cr
&& && && && \cr
&& && && && \cr
&& && && && \cr
&& && && && \cr
S^3/D^*_8&& * &&O^*&&0&&\pi_k(S^3)\cr
S^3/D^*_{4n}&& * &&D^*_{8n}&&0&&\pi_k(S^3)\cr
S^3/T^*&& ? &&O^*&&0&&\pi_k(S^3)\cr
S^3/O^*&& w &&O^*&&0&&\pi_k(S^3)\cr
S^3/I^*&& ? &&I^*&&0&&\pi_k(S^3)\cr
&& && && && \cr
&& && && && \cr
&& && && && \cr
S^3/D^*_8\times Z_p&& * &&Z_2\times O^*&&Z&&\pi_k(S^3\times S^3)\cr
S^3/D^*_{4n}\times Z_p&& * &&Z_2\times D^*_{8n}&&Z&&\pi_k(S^3\times S^3)\cr
S^3/T^*\times Z_p&& ? &&Z_2\times O^*&&Z&&\pi_k(S^3\times S^3)\cr
S^3/O^*\times Z_p&& w &&Z_2\times O^*&&Z&&\pi_k(S^3\times S^3)\cr
S^3/I^*\times Z_p&& ? &&Z_2\times I^*&&Z&&\pi_k(S^3\times S^3)\cr
S^3/D'_{2^k(2n+1)}\times Z_p&& * &&Z_2\times D^*_{8(2n+1)}&&Z&&\pi_k(S^3 \times
S^3)\cr
S^3/T'_{8\cdot 3^k}\times Z_p&& ? &&O^*&&Z&&\pi_k(S^3\times S^3)\cr
&& && && \cr
&& && && \cr
&& && && \cr
L(p,q_1)&& w &&Z_2&&Z&&\pi_k(S^3)\cr
L(p,q_2)&& w,* &&Z_2\times Z_2&&Z\times Z&&\pi_k(S^3\times S^3)\cr
L(p,q_3)&& w &&Z_4&&Z\times Z&&\pi_k(S^3\times S^3)\cr
L(p,q_4)&& w &&Z_2&&Z\times Z&&\pi_k(S^3\times S^3)\cr
RP^3&& * &&1&&0&&0\cr
S^3&& * &&1&&0&&0\cr
&& && && && \cr
&& && && && \cr
&& && && && \cr
S^2\times S^1&& &&Z_2\times Z_2&&Z&&\pi_k(S^3\times S^2)\cr
&& && && && \cr
&& && && && \cr
&& && && && \cr
K(\pi, 1)_{\hbox{sl}}&& &&\hbox{Aut}^{Z_2}_+(\pi_1)&&0&&
                                            \pi_k(S^3)\cr}}
$$

The table summarizes the results for the homotopy groups of the configuration
spaces \Q(\S). The last column comprises all homotopy groups higher than the
second, i.e., $k\geq 3$. The table is divided into five horizontal blocks
of which the first three represent the spherical space forms, the fourth the
single handle-manifold (wormhole), and the fifth the sufficiently large
$K(\pi,1)$ primes. The calculation for the last two blocks is almost trivial
in view of the strong results obtained in [Ha3] and [Ha1] respectively.
${\rm Aut}^{Z_2}_+$ in the last block denotes a $Z_2$ extension (according
to (2.6) imposed by spinoriality) of the orientation preserving automorphisms
of $\pi_1$.

The calculations for the spherical space forms depend on the validity of the
Hatcher conjecture (HC), which is not known to hold in all cases. The
calculations of $\pi_1(\Q)$, however, depend only on a weak implication
thereof, namely $\pi_0(D_F)\cong \pi_0(Isom)$. Since there are cases where
this weak form of HC but not HC itself is known to hold, we indicated their
status of validity separately in the second column (compare Appendix 2,
Theorems A1 and A2). Here, an asterisk, (*), denotes the validity of HC,
a $w$ the validity of the weak form only, and the question mark, (?), that
no such result is known to us. Assuming ($w$), the groups $\pi_1(\Q(S^3/G))$
were first calculated in [Wi]. Subsequently the table was completed to the
present form in [Gi1].

The first block contains the homogeneous spherical primes with non-cyclic
fundamental group, the second the non-homogeneous ones. Here, the order of the
additional cyclic group, $p$, has to be coprime to the order of the group the
$Z_p$ is multiplied with, and $\geq 2$ in the first five, and $\geq 1$ in the
remaining two cases. The third block contains all the spherical primes with
cyclic fundamental group $Z_p$, otherwise known as the lens spaces $L(p,q)$,
where $q$ has to be coprime to $p\geq 2$. Here, $q_1$ stands for $q=\pm 1$
mod $p$, $q_2$ for $q\not=\pm1$ mod $p$ and $q^2=1$ mod $p$, $q_3$ for
$q^2=-1$ mod $p$, and $q_4$ for the remaining cases.
Amongst all $L(p,q_2)$ are those of the form $L(4n,2n-1),n\geq 2$. For those
the (*) is valid in the second column and ($w$) for all others. Finally,
$RP^3$ and $S^3$ are listed separately. Together with $L(p,q_1)$ they comprise
the homogeneous spaces in the third block, and, taken together with the first
block, all the homogeneous spherical primes in the list.

\subsection{Some Observations}

In standard canonical quantization one regards the wave function as a section
of a possibly non-trivial complex line bundle over some configuration space.
For not simply connected configuration spaces, this induces an action of the
fundamental group on sections via some 1-dimensional representations and thus
also on the space of quantum states. In the canonical quantization approach to
quantum gravity (as in other field theories) one just tries to repeat the same
construction, using the classical configuration space as domain space for the
state functional.

{\baselineskip=20truept\parskip=0pt{\eightpoint
Since this space is infinite dimensional, the construction of a
quantum state space (Hilbert space of sections) is technically
much more difficult and overshadowed by the lack of appropriate
measures. This points towards the necessity to use a kind of
distributional-dual as domain space, on which such measures are
available. This is not particular to theories of gravity and
renders the availability and construction of the Schr\"odinger
representation a non-trivial technical problem in any field theory.
In our case, for example, it is easy to see that the group $D_F$
does not act freely on the distributional dual of the space of
Riemannian metrics, thus rendering an analogous structure to (1.3)
and hence the given derivation of (1.5) impossible. But we feel
that a better technical definition and understanding of canonical
quantisation in field theory is required first to attack these
issues profitably. The presently given standard arguments are
therefore necessarily heuristic in nature, as already stated in
the Introduction.

As far as the fundamental group is concerned, it is sometimes
alternatively argued directly using a presumed action of the
zeroth homotopy group of the group of gauge
transformations (diffeomorphisms) on the space of quantum states,
without trying to use the auxiliary structure of the classical
configuration space for its explicit construction. The (pretended)
Schr\"odinger representation might then be useful in trying to
visualize this action in terms of physical operations (e.g. exchanges,
rotations etc.) on prefered states.
\par}}

Applied to the case at hand, the possible fundamental groups are
determined by (1.8) with those occuring in the table above.
In this context one would say that a spinorial manifold $\S$ admits
{\sl abelian} spinorial states, if $\pi_1(\Q(\SC))$ allows for one dimensional
representations which represent the $Z_2$ generated by a $2\pi$-rotation
parallel to an asymptotic sphere non-trivially. States transforming
non-trivially under such a rotation where first considered in [Fr-So].
Their assertion was that if a single state existed that was not left
invariant by this $Z_2$, its antisymmetric combination with the $Z_2$
transformed one would be a state that changed sign under that $Z_2$.
They, however, did not address the question of whether such a state carried
any one-dimensional representation of $\pi_1(\Q(\SC))$.
Here, we make the following

\proclaim Observation 1.
No 3-manifold, $\SC$, whose prime decomposition consists entirely
of primes taken from the upper four horizontal blocks in our table
allows for abelian spinorial states.\par

\noindent{\bf Proof}. Clearly we assume the presence of at least one
spinorial prime, since the statement is trivial otherwise. As shown in the
proof for Corollary 1, the rotation parallel to the sphere at $\infty$
is isotopic to rotations parallel to each connecting sphere for
the primes. This generates an extending $Z_2$ subgroup (see 2.6) in each
factor in (1.8) that corresponds to a spinorial prime. $\SC$ thus allows for
abelian spinorial states, if and only if at least one of the spinorial $\SC_i$
does.
However, none of the groups $O^*, D^*_{8m}\,m\geq 1, I^*$ (see table) has a
1-dimensional representations that represent the extending $Z_2$ of their
$SO(3)$
counterparts non-trivially.

{\baselineskip=20truept{\eightpoint
The latter fact can be proven from the presentations
$$\eqalign{
D^*_{8n}&=\{x,y\,\vert\, x^2=(xy)^2=y^{2n}\}\cr
O^*     &=\{x,y\,\vert\, x^2=(xy)^3=y^4;\,x^4=1\}   \cr
I^*     &=\{x,y\,\vert\, x^2=(xy)^3=y^5;\,x^4=1\}\,,\cr}
$$
from which their abelianizations are readily determined to be $Z_2\times Z_2$
(generated by $x$ and $y$, taken as abelianized generators), $Z_2$
(generated by $x$) and 1 respectively. Since in each case it is $x^2$ that
generates the extending $Z_2$, the result follows. Note that e.g.
$D^*_{4(2n+1)}$ would have the desired representation since its abelianisation
is $Z_4$, generated by $x$. It just does not happen to occur as a $\pi_1(Q)$.
Note also that $x^2$ is always a central element. For this reason
one can define an action of $2\pi$-rotations on general wavefunctions not
carrying one-dimensional representations (see e.g. [Gi2])$\bullet$
\par}

We do not know whether the result can easily be extended to include general
$K(\pi,1)$-primes as well. We explicitly checked, however, that e.g. the
three-torus as well does not allow for abelian spinorial states. But the
situation clearly changes if one allows for higher dimensional representations.
For example, the group $O$ has five irreducible representations of
dimensions 1, 1, 2, 3 and 3, whereas $O^*$ has in addition three more
of dimensions 2, 2 and 4, all of which represent the extending $Z_2$
non-trivially. If one allows for multiple component wavefunctions
(sections in higher dimensional complex vector bundles), spinorial
states can be incorporated, albeit in this case one can not incorporate
the whole group as transformation group [An][Gi2]. Let us note at this
point that higher dimensional representations automatically appear in
carrying through the standard quantum-mechanical formalism in presence of a
discrete gauge group. Excluding them means that one a priori excludes
potentially interesting sectors [Gi2].

Pushing the original setting a bit further, we remark that
the possible, inequivalent line bundles with connection over $\Q$ are
classified by $H_1(\Q,Z)\oplus FH_2(\Q,Z)$ (see e.g.[Wo]), where $F$
denotes the free part. It is also convenient to split $H_1$ into
a free-, and a torsion part (denoted by T). The free part of $H_1$ alone
then accounts for the different flat connetions with unchanged bundle
topology, whereas $TH_1(\Q,Z)\oplus FH_2(\Q,Z)\cong H^2(\Q,Z)$ labels
the topologically inequivalent bundles. Only the latter define different
sectors for the quantum theory, and may e.g. show up as a non-trivial
spectral flow for the Dirac operator [Al-Nel]. Whereas
for non-trivial $FH_2$ one cannot make any statements without explicitly
analysing the dynamics as to whether it actually makes use of a
non-trivial class, a trivial $FH_2$ clearly excludes such possibilities
from the beginning. It is therefore interesting to see whethere there are
$\SC$'s whose associated $\Q(\SC)$'s have trivial $FH_2(\Q,Z)$. From the
table  and (1.8) we see that this is generaly not the case. However, we
have the

\proclaim Observation 2. Let $\SC$ be a homogeneous spherical prime
of non-cyclic fundamental group and $\Q=\Q(\SC)$ its associated
configuration space; then $FH_2(\Q,Z)=0$. \par

\noindent{\bf Proof.} From the table we infer that for the $\SC$ in question
the associated $\Q(\SC)$ have finite universal cover of which the first two
homotopy groups are trivial. A standard spectral sequence argument, which
we suppress at this point, then shows that $H_2(\Q,Q)$ (rational coefficients)
is trivial, which is equivalent to the statement made $\bullet$  \par

\noindent Taken together with Observation 1 this excludes a possibly
conjectured connection between spinoriality and the possibility of
non-trivial spectral flows.

A very special case is that of $\SC=RP^3$. Here we have immediately

\proclaim Observation 3. $RP^3$ is the unique, non-trivial prime
3-manifold on our table whose configuration space has alltogether trivial
homotopy (and hence homology) groups. \par

{}From (1.8) one thus infers that a multi-$RP^3$ manifold, $\SC$, therefore
receives all the non-trivial topology of $\Q(\SC)$ from the factor
$\Omega\C$. A more detailed study of $\C$ shows [He-McCu] that in this case
$\pi_*(\Q(\SC))$ contains as subgroups those for the configuration space
of $n$ identical objects in $R^3$ (if $\SC$ is the connected sum of $n$
$RP^3$'s). Its fundamental group is $P_n$, the permutation group of $n$
objects, and the higher ones are given in ref. [Fa-Neu]. This particular
example has been used to discuss, and rule out, the obvious version of
a spin-statistics relation, as generally discussed in [So]. However,
that such a correlation cannot exist is a general consequence of the
product formula (1.8), which explicitly uncorrelates the representation
chosen for rotations and exchanges, irrespectively of the different
definitions of the latter [An].

\vfill\eject

\beginsection{Appendix 1}

In this appendix we show how to prove non-spinoriality for $\SC=L(p,q)$ and
$\SC=S^1\times S^2$. To this end we pick a curve $\alpha:\,[0,1]\rightarrow
SO(3)$ that generates $Z_2=\pi_1(SO(3),id)$, find a covering curve in
$D_{\infty}$, starting at the identity, and show that it ends in the identity
component $D_F^0$ of $D_F$ (we refer to (2.3)). We may choose
$\alpha(s)=\exp(2\pi s\hat z)$ and denote the corresponding linear map in
$R^3$ by $R_z[2\pi s]$, or $R_z[\varphi]$ for general angles~$\varphi$.

Let now $\sigma:\,R^3\supset B_2\rightarrow \SC$ be an embedding of
$B_2=\{x\in R^3\,/\, \| x\|\leq 2\}$ into $\SC$. We let $r$
denote the distance from the origin of $B_2$ and set:
$$\eqalign{
D_2&=Image\, \sigma\big\vert_{r\leq 2}\,,\quad
D_1=Image\, \sigma\big\vert_{r\leq 1}\,, \quad
T=Image\, \sigma\big\vert_{1\leq r\leq 2}      \cr
S_2&=Image\, \sigma\big\vert_{r=2}\,,\quad
S_1=Image\, \sigma\big\vert_{r=1}\,,\quad
\infty=Image\, \sigma\big\vert_{r=0}         \cr}
\eqno{(A1)}
$$
On $B_2$ we can define a path of homeomorphisms, $\rho_s$, by
$$
\rho_s=\cases{R_z[2\pi s(2-r)] & for $2\geq r\geq 1$   \cr
              R_z[2\pi s]      & for $r\leq 1$ ,       \cr}
\eqno{(A2)}
$$
which then defines a path of homeomorphisms, $R_s$, of $\SC$, by setting
$$
R_s=\cases{\sigma\circ\rho_s\circ\sigma^{-1} & on $D_2$      \cr
           id                                & on $\SC-{\displaystyle
                                               {\mathop{D}^{\circ}}_2}$ .\cr}
\eqno{(A3)}
$$
This map is not differentiable on $S_2$ and $S_1$, but it may easily be
smoothed by modifying it in arbitrarily small collar-neighbourhoods of these
two spheres. We imagine this being done but without giving the details
here, since in order to caculate the projection map $\tilde p$ in (2.3)
we only need differentiability in a neigbourhood of $\infty$. $R_s$
``rigidly" rotates $D_1$ by an angle $2\pi s$ about the $z$-axis by
progressing ``rigid" rotations of the spheres $r=const.$ within $T$.
For $s=1$ we have $R_s\big\vert_{D_1}=id$ so that we call $R_1$ a rotation
parallel to $S_1$ and $S_2$ (see e.g. [He][L]).
Each $R_s$ fixes $\infty$ and projects to $\alpha(s)$ via $\tilde p$.
$R_1$ is in $D_F$ since it fixes a disc ($D_1$) containing $\infty$.
In order to show that $R_1$ is in $D_F^0$, we now explicitly construct a
path, $K_s$, from $id$ to $R_1$ within $D_F$. As above, it is sufficient to
construct a path of homeomorphisms rather than diffeomorphisms that fix $D_1$
and which we then imagine to be smoothed appropriately. The details are
irrelevant for us. For the construction it is convenient to represent the
spaces $L(p,q)$ and $S^1\times S^2$ by the following fundamental domains:

\noindent{\bf $L(p,q)$:} Take a solid ball $\| x\|\leq 3$ in R$^3$,
and identify the 2-dimensional sectors $s_k:\,{2\pi\over p}(k-1)\leq\phi
\leq{2\pi\over p}k$ on the upper hemisphere with the sectors $s_{k+q}$
on the lower hemisphere, by first reflecting them on the equatorial plane
($z=0$), followed by a rotation about the $z$-axis. We take
$T=\{x\in R^3\,/\,1\leq\| x\|\leq 2\}$, so that $\infty$ corresponds
to $x=0$.

\noindent{\bf $S^1\times S^2$:} Take a solid sphericall shell
$1\leq\vert x\vert \leq 6$ in $R^3$, and identify the inner and outer
2-sphere boundaries radially (i.e. points of equal polar angles are
identified). We take $T=\{x\in R^3\,/\,1\leq \vert x-(0,0,3)\vert\leq 2\}$
so that $\infty$ corresponds to $(0,0,3)$.

The crucial observation is that the $SO(3)$- rotation, $R_z[\varphi]$,
applied to these domains is compatible with the boundary identifications,
and, therefore, defines a homeomorphism of the manifolds in question.
But then it is obvious how to reach $R_1$ by a path of homeomorphisms that
fix $D_1$: instead of rotating $D_1$ against $S_2$, we rotate $\SC-D_2$
and with it $S_2$ against $S_1$ by just the negative amount. That is
($r$ still denotes the distance from $\infty$ and $\sigma$ is the identity
since we work within $R^3$)
$$
K_s=\cases{id                & on $D_1:\,(r\leq 1)$     \cr
           R_z[-2\pi s(r-1)] & on $D_2-{\displaystyle{\mathop{D}^{\circ}}_1}
                                   :\,(1\leq r\leq 2)$  \cr
           R_z[-2\pi s]      & on $\SC-{\displaystyle{\mathop{D}^{\circ}}_2}
                                   :\, r\geq 2$         \cr}
\eqno{(A4)}
$$
so that
$$
K_1=\cases{id                & on $D_1$         \cr
           R_z[2\pi(1-r)]    & on $D_2-{\displaystyle{\mathop{D}^{\circ}}_1}$
                                                                          \cr
           R_z[-2\pi]        & on $\SC-{\displaystyle{\mathop{D}^{\circ}}_2}$
                                                                          \cr}
\eqno{(A5)}
$$
which, by $R_z[\varphi + n\,2\pi]=R_z[\varphi]$, is equal to $R_1$.

{\baselineskip=20truept{\eightpoint
Although we work with oriented manifolds only, let us note for
completeness that from the fundamental domain we used to construct
$S^1\times S^2$, we can also construct $S^2\tilde{\times} S^2$, the
unique non-orientable 2-sphere bundle over $S^1$. For this we just
identify the inner and outer 2-sphere boundaries via the antipodal
map (in standard polar angles:
$(\theta, \phi) \mapsto (\pi-\theta, \phi+\pi)$), rather than using
the identity. The rotations $R_z[\varphi]$ are then still compatible
with the  boundary identifications, hence proving non-spinoriality for
$S^1\tilde{\times}S^2$.\par}}

\vfill\eject

\beginsection{Appendix 2}

In this appendix we collect some of the results on the diffeomorphism
group of 3-manifolds which were of relevance in our investigations.

\proclaim Theorem A1. For spherical primes ($S^3/G$), the handle
($S^1\times S^2$), sufficiently large $K(\pi,1)$'s ($K(\pi,1)_{sl}$) and
most of the non-sufficiently large $K(\pi,1)$'s which are Seifert,
two diffeomorphisms are isotopic if and only if they are homotopic.

\noindent{\bf Proof.} For the spherical primes these statemants are proven
in [Ho-R] (Lens spaces $L(p,q)$), [Bir-R]
(octahedral spaces $S^3/O^*\times Z_p$), [Asa] or [R]
(prism and generalized prism spaces $S^3/D^*_{4m}\times Z_p$,
$S^3/D'_{2^k\cdot (2n+1)}\times Z_p$ and [Boi-O] (icosahedral,
tetrahedral and generalized tetrahedral $S^3/I^*\times Z_p$,
$S^3/T^*\times Z_p$, $S^3/T'_{8\cdot 3^k}\times Z_p$). For the handle this
is proven in [Gl], for the $K(\pi,1)_{sl}$'s in [Wa] and for the
Seifert non-sufficiently large $K(\pi,1)$'s in [Sc].
$\bullet$

\proclaim Theorem A2. For all sperical primes but the icosahedral, tetrahedral
and generalized tetrahedral spaces one has the isomorphism
$\pi_0(D)\cong\pi_0(Isom)$, where $Isom$ denotes the space
of orientation preserving isometries.

\noindent{\bf Proof.} $\pi_0(D)$ has been calculated in the refernces just
given, except [Boi-O]. The calculations for $\pi_0(Isom)$ may be
found in [Wi].
$\bullet$

For the 3-sphere it is proven in [Ha4] that there is a homotopy equivalence
$D(S^3)\sim Isom (S^3)$. As generalisation it has been conjectured in [Ha2]
that there is a homotopy equivalence for all spherical primes
(Hatcher conjecture (HC)): $D(S^3/G)\sim Isom(S^3/G)$.
The efforts to prove this are so far summarized in the following

\proclaim Theorem A3. HC holds for the real projective space $RP^3$, the
lens spaces $L(4k,2k-1)$, where $k>1$, and the prism and generalized prism
spaces $S^3/D^*_4\times Z_p$ and $S^3/D'_{2^k\cdot (2n+1)}\times Z_p$
respectively.\par

\noindent{\bf Proof.} The assertion for $RP^3$ is made in [Ha2].
For $S^3/D^*_{4n}\times Z_p$, $p>1$, and
$S^3/D'_{2^k\cdot (2n+1)}\times Z_p$, $p\geq 1$, this was proven in [I1][I2]
and for $S^3/D^*_{4n}$ and $L(4k,2k-1)$ $k>1$ in [McCu-R]
$\bullet$\par

Similar results are known for the handle  and the $K(\pi,1)_{sl}$'s:

\proclaim Theorem A4. The group of diffeomorphisms of $S^2\times S^1$ has the
homotopy type of $O(2)\times O(3)\times\Omega O(3)$, where $\Omega(\cdot)$
denotes the loop space of ($\cdot$). \par

\noindent{\bf Proof.} This is proven in [Ha3]. If one considers
orientation preserving diffeomorphisms only, one may write:
$D(S^2\times S^1)\cong Z_2\times S^1\times SO(3)\times\Omega SO(3)$
$\bullet$\par

\proclaim Theorem A5. Let $\SC$ be an orientable, sufficiently large
$K(\pi,1)$ prime of fundamental group $G$. Its diffeomorphism group, $Diff$,
has homotopy groups\hfill\break
$\pi_0(Diff)\cong Out(G)$, $\pi_1(Diff)\cong centre\,G$
and $\pi_k(Diff)=0\quad \forall k\geq 2$.\par

\noindent{\bf Proof.} The proof is given in [Ha1] for the larger
class of $P^2$-irreducible (irreducible which contain no two-sided $RP^2$),
sufficiently large $K(\pi,1)$'s. If one restricts to orientation preserving
diffeomorphisms, one has $\pi_0(D)\cong Out_+(G)$, the outer
automorphisms which respect the orientation homomorphism (a $Z_2$-quotient
of $Out(G)$ if $\SC$ allows for orientation reversing diffeomorphisms)
$\bullet$

\vfill\eject

\beginsection{References}

\item{[Al-Nel] } Alvarez-Gaum\'e, L., Nelson, P.: Hamiltonian
                 interpretation of anomalies. Comm. Math. Phys.,
                 {\bf 99}, 103-114 (1985).

\item{[An] }     Aneziris, C., Balachandran, A.P., Bourdeau, M., Jo, S.,
                 Ramadas, T.R., Sorkin, R.: Aspects of spin and statistics
                 in generally covariant theories. Int. Jour. Mod. Phys. A,
                 {\bf 4}, 5459-5510 (1989).

\item{[Asa] }    Asano, K.: Homeomorphisms of prism manifolds. Yokohama
                 Math. Jour., {\bf 26}, 19-25 (1978).

\item{[Ash] }    Ashtekar, A.: Lectures on non-perturbative canonical
                 gravity. Advanced Series in Astrophysics and Cosmology
                 Vol. 6. World Scientific, Singapore, New Jersey, London,
                 Hong Kong 1991.

\item{[Bir-R] }  Birman, J.S., Rubinstein, J.H.: Homeotopy groups of some
                 non-Haken manifolds. Proc. London math. Soc., {\bf 49},
                 517-536 (1984).

\item{[Bou] }    Bourguignon, J-P.: Une stratification de l'espace des
                 structures Riemanniennes. Comp. Math., {\bf 30}, 1-41
                 (1975).

\item{[Boi-O] }  Boileau, M., Otal, J-P.: Groupe des diffeotopies de
                 certaines vari\'et\'es de Seifert. C.R. Acad. Sc. Paris,
                 {\bf 303}, Ser. 1, 19-22 (1986).

\item{[C] }      Cerf, J.: groupes d'automorphismes et groupes de
                 diff\'eomorphismes des vari\'et\'es compactes de
                 dimension 3. Bull. Soc. Math. France, {\bf 87}, 319-329
                 (1959).

\item{[DeW] }    DeWitt, B.: Spacetime as a sheaf of geodesics in superspace.
                 In: Relativity, Proceedings of the Relativity conference in
                 the Midwest, held at Cincinnati, Ohio, June 2-6, 1969.
                 Ed. Carmeli, M., Fickler, S.I., Witten, L. Plenum Press,
                 New York - London, pp 359-374 (1970).

\item{[Fa-Neu] } Fadell, E., Neuwirth, L.: Configuration spaces. Math.
                 Scand., {\bf 10}, 111-118 (1962).

\item{[Fi1] }    Fisher, A.E.: The theory of superspace. In: Relativity,
                 Proceedings of the Relativity conference in the Midwest,
                 held at Cincinnati, Ohio, June 2-6, 1969. Ed. Carmeli,
                 M., Fickler, S.I., Witten, L. Plenum Press, New York -
                 London, pp 303-357 (1970).

\item{[Fi2] }    Fisher, A.E.: Resolving the singularities in the space
                 of Riemannian geometries. Jour. Math. Phys., {\bf 27},
                 718-738 (1986).

\item{[Fr-So] }  Friedman, J., Sorkin, R.: Spin ${1\over 2}$ from Gravity.
                 Phys. Rev. Lett., {\bf 44}, 1100-1103 (1980).

\item{[Fr-Wi] }  Friedman, J., Witt, D.: Homotopy is not isotopy for
                 homeomorphisms of 3-manifolds. Topology, {\bf 25}, 35-44
                 (1986).

\item{[Gi1] }    Giulini, D.: 3-manifolds in canonical quantum gravity.
                 Thesis, University of Cambridge (GB), (1990).

\item{[Gi2] }    Giulini, D.: Quantum mechanics on spaces with non-abelian
                 finite fundamental group. In preparation.

\item{[Gl] }     Gluck, H.: The embedding of two-spheres in the four-sphere.
                 Bull. Amer. Math. Soc., {\bf 67}, 586-589 (1961).

\item{[Ha1] }    Hatcher, A.: Homeomorphisms of sufficiently large
                 $P^2-$irreducible 3-manifolds. Topology, {\bf 15}, 343-347
                 (1976).

\item{[Ha2] }    Hatcher, A.: Linearisation in 3-dimensional topology.
                 Proceedings of the international congress of mathematicians,
                 Helsinki, 1978, Vol. {\bf 1}, 463-468. Ed. Olli Lehto.

\item{[Ha3] }    Hatcher, A.: On the diffeomorphism group of $S^2\times S^2$.
                 Proc. Amer. Math. Soc., {\bf 83}, 427-430 (1981).

\item{[Ha4] }    Hatcher, A.: A proof of the Smale conjecture,
                 $Diff(S^3)\cong O(4)$. Annals of Math., {\bf 117}, 553-607
                 (1983).

\item{[He] }     Hendriks, H.: Application de la th\'eorie d'obstruction
                 en dimension 3. Bull. Soc. Math. France, suppl\'ement, memoire
                 {\bf 53}, 81-195 (1977).

\item{[He-L] }   Hendriks, H., Laudenbach, F.: Diffeomorphismes des sommes
                 connexes en dimension trios. Topology, {\bf 23}, 423-443
                 (1984).

\item{[He-C] }   Hendriks, H., McCullough, D.: On the diffeomorphism group
                 of a reducible 3-manifold. Topology and its Application,
                 {\bf 26}, 25-31 (1987).

\item{[Ho-R] }   Hodgson, C., Rubinstein, J.H.: Involutions and isotopies
                 of lens spaces. In: Knot theory and manifolds, ed. by
                 D. Rolfsen. Proceedings, Vancover 1983. Lecture Notes in
                 Mathematics, {\bf 1144}, 60-96 (1985), Springer Verlag.

\item{[I1] }     Ivanov, N.V.: Homotopy of spaces of automorphisms of some
                 three-dimensional manifolds. Sov. Math. Dokl., {\bf 20},
                 47-50 (1979).

\item{[I2] }     Ivanov, N.V.: Homotopy of spaces of diffeomorphisms of some
                 three-dimensional manifolds. Jour. Sov. Math., {\bf 26},
                 1646-1664 (1984).

\item{[L] }      Laudenbach, F.: Topologie de la dimension trois; homotopie
                 et isotopie. Asterisque, {\bf 12}, 1-137 (1974).

\item{[McCa] }   McCarty, G.S.(Jr.): Homeotopy groups. Trans. Amer. Math.
                 Soc., {\bf 106}, 293-303 (1963).

\item{[McCu] }   McCullough, D.: Mappings of reducible manifolds.
                 Geometric and Algebraic Topology; Banach center publications,
                 {\bf 18}, 61-76 (1986). PWN - Polish scientific publishers,
                 Warszawa.

\item{[McCu-R] } McCullough, D., Rubinstein, J.H.: The generalized Smale
                 conjecture for 3-manifolds with genus 2 one-sided Heegaard
                 splittings. University of Oklahoma preprint (1990).

\item{[McD] }    McDuff, D.: The Lattice of Normal Subgroups of the
                 Group of Diffeomorphisms or Homeomorphisms of an Open
                 Manifold. Jour. London Math. Soc., {\bf 18} (2. Serie),
                 353-364 (1978)

\item{[Mi] }     Michor, P.W.: Manifolds of Differentiable Mappings. Shiva
                 Mathematical Series 3, Shiva Publishing Limited, UK, 1980

\item{[R] }      Rubinstein, J.H.: On 3-manifolds that have finite fundamental
                 group and contain Klein bottles. Trans. Amer. Math. Soc.,
                 {\bf 251}, 129-137 (1979).

\item{[SC] }     Scott, P.: Homotopy implies isotopy for some Seifert fibre
                 spaces. Topology, {\bf 22}, 341-351 (1985).

\item{[Sm] }     Smale, S.: Diffeomorphisms of the 2-sphere. Proc. Amer.
                 Math. Soc., {\bf 10}, 621-626 (1959).

\item{[So] }     Sorkin, R.: A general relation between kink-exchange and
                 kink-rotation. Comm. Math. Phys., {\bf 115}, 421-434 (1988).

\item{[Wa] }     Waldhausen, F.: On irreducible 3-manifolds which are
                 sufficiently large. Annals of Math., {\bf 57}, 56-88 (1968).

\item{[Wi] }     Witt, D.: Symmetry groups of state vectors in canonical
                 quantum gravity. Jour. Math. Phys., {\bf 27}, 573-592 (1986).

\item{[Wo] }     Woodhouse, N.: Geometric Quantisation. Oxford Mathematical
                 Monographs, Claredon Press, Oxford 1980.

}
\end